\documentclass[24pt]{article}
\usepackage{amssymb}
\usepackage{epsfig,caption,graphicx}
\usepackage{pstricks}
\usepackage{amssymb,amsmath}
\usepackage{multido}
\usepackage{array}
\begin{document}
%\draft
\def\be{\begin{equation}}
\def\ee{\end{equation}}
\def\ba{\begin{eqnarray}} 
\def\ea{\end{eqnarray}}
\def\nn{\nonumber}

\newcommand{\bbf}{\mathbf}   
\newcommand{\rrm}{\mathrm}
\title{Structure of the interaction and energy transfer between an open quantum system and its environment\\}

\author{Tarek Khalil$^{a}$
\footnote{E-mail address: tkhalil@ul.edu.lb}\\ 
and\\
Jean Richert$^{b}$
\footnote{E-mail address: j.mc.richert@gmail.com}\\ 
$^{a}$ Department of Physics, Faculty of Sciences(V),\\
Lebanese University, Nabatieh,
Lebanon\\
$^{b}$ Institut de Physique, Universit\'e de Strasbourg,\\
3, rue de l'Universit\'e, 67084 Strasbourg Cedex, 
France}
\date{\today}   
\maketitle 
\begin{abstract}
Due to the coupling of a quantum  system to its environment energy can be transfered between the two subsystems in both directions. In the present study we consider this process in a general framework for interactions with different properties and show how these properties govern the exchange.   
\end{abstract} 
\maketitle

\section{Introduction}

Quantum systems are generally never completely isolated but interact with an environment with which they may exchange energy and other physical observable quantities. Their properties are naturally affected by their coupling to the external world. The understanding and control of the influence of an environment on a given physical system is of crucial importance in different fields of physics and in technological applications and led to a large amount of investigations ~\cite{reb,aba,kos,gol,cor}.
Among the quantities of interest energy exchange between a system and its environment is of prime importance. The total system composed of the considered system and its environment is closed. It conserves the physical observables, in particular the total energy it contains. However, the existence of an interaction between its two parts leads to possible exchanges between them, energy and other observables can be transfered between them in both directions. In the present work we study the conditions under which this tranfer occurs.

In order to investigate the process a number of different approaches have been developed
\cite{bel,bag,fl1,fl2,esp}. They make use of the cumulant method as well as other developments ~\cite{car,sch,def}. In the following we shall use the cumulant method in order to work out the energy exchange and the speed at which this exchange takes place~\cite{gua}.

In previous studies ~\cite{kr1,kr2,kr3} we defined criteria which allow to classify open quantum systems with respect to their behaviour in the presence of an environment. In order to do so we used a general formulation which relies on the examination of the properties of the density operator of the system and its environment. The dynamical behaviour of this operator is governed by the structure of the Hamiltonians of the system, its environment and the coupling Hamiltonian which acts between them. The method introduces a general form of the total density operator and avoids the determination of explicit solutions of the equation of motion which governs the time evolution of the system. We follow this formalism in order to work out different cases of physical interest concerning the energy exchange process.\\    

The analysis is presented in the following way. In section 2 we define the energy exchange and its rate starting from the characteristic function which generates these quantities in terms of its first and second moment. Section 3 introduces a general formal expression of the density operator at the initial time and the structure of the total Hamiltonian of the system, the environment and their interaction. The expressions of the energy exchange and the exchange rate are explicitly written out. In section 4 we analyze and work out two different cases. They exemplify the role of the interaction  between the system and the environment which may or may not induce the time divisibility property of the open system. Conclusions are drawn in section 5.

%\newpage
 
\section{Energy transfer between S and E: the cumulant approach}

Consider a system $S$ coupled to an environment $E$. The time dependent density operator of the total system is $\hat \rho_{SE}(t)$ and $S$ and $E$ are coupled by an interaction $\hat H_{SE}$. The interaction may generate an energy exchange between the two parts. This quantity can be worked out by means of the cumulant method ~\cite{gua} which was developed in a series of works, first for closed driven systems ~\cite{esp}, later extended to open systems, see f.i.~\cite{fl1,fl2} and further approaches quoted in ~\cite{gua}.

Define the modified density operator

\ba
\rho^{\sim}_{SE}(t,0)=Tr_{E}[{\hat U_{\eta/2}(t,0)\hat \rho_{SE}(0)\hat U^{+}_{-\eta/2}(t,0)}]
\label{eq1} 
\ea 
where $Tr_{E}$ is the trace over the states in $E$ space and
\ba
\hat U_{\eta}(t,0)=\exp(+i\eta \hat H_{SE})\hat U(t,0)\exp(-i\eta \hat H_{SE})
\label{eq2} 
\ea
Here $\hat U(t,0)$ is the time evolution operator of the interacting total system $S+E$. 

The characteristic function obtained from the generating function reads 
\ba
\chi^{(\eta)}(t)=Tr_{S}(\rho^{\sim}_{SE}(t,0))
\label{eq3} 
\ea

From $\chi^{(\eta)}(t)$ one derives the energy exchange between the environment and the system 

\ba
\Delta E(t)=\frac{d \chi^{(\eta)}(t)}{d (i\eta)}_{| \eta=0}
\label{eq4} 
\ea
The speed at which the energy flows between the system and its environment is given by 

\ba
V_{E}(t)=\frac{\partial \dot \chi^{(\eta)}(t)}{\partial (i\eta)}_{|\eta=0}
\label{eq5} 
\ea
where $\dot \chi^{(\eta)}(t)$ is the time derivative of $\chi^{\eta}(t)$. If the energy flows from the sytem to the environment $V_{E}(t)$ is positive and negative if the flow is reversed.

%\newpage

\section{Energy flow and speed of energy exchange}

\subsection{The density operator}

At time $t=0$ we choose the system to stay in a mixed state

\ba
|\psi(0)\rangle=\sum_{{i_k}}c_{i_k}|i_k\rangle
\label{eq6} 
\ea
where the normalized states $\{|i_{k}\rangle\}$ are eigenstates of the Hamiltonian $\hat H_{S}$. The environment is described in terms of its density matrix chosen as 
$\{ |\alpha\rangle d_{\alpha,\alpha}\langle \alpha|\}$ where $\{d_{\alpha,\alpha}\}$ are the statistical weights of the density matrix in a diagonal basis of states $\{\alpha\}$.

Given these bases of states in $S$ space and in $E$ space the density operator of the total system at
time $t=0$ is written as~\cite{buz}

\ba
\hat \rho_{SE}(0)=\hat \rho_{S}(0) \otimes \hat \rho_{E}(0)
\notag\\
%\hat \rho_{S}(0)=\sum_{(i_1,i_2)}c_{i_1}|i_1\rangle \i_{2}}|c^{*}_{i_2}
\hat \rho_{S}(0)=\sum_{k,l}|i_{k}\rangle c_{i_{k}}c^{*}_{i_{l}}\langle i_{l}|
\notag\\
%\hat \rho_{E}(0)=\sum_{\alpha_1,\alpha_2}d_{\alpha_1,\alpha_2}|\alpha_1\rangle \langle \alpha_{2}|
\hat \rho_{E}(0)=\sum_{\alpha}|\alpha \rangle d_{\alpha,\alpha} \langle \alpha|
\label{eq7} 
\ea
The density operator $\hat \rho_{SE}$ describes a system in the absence of an interaction 
$\hat H_{SE}$ at $t=0$, hence in the absence of an initial entanglement between $S$ and $E$. 

The total Hamiltonian $\hat H$ reads

\ba
\hat H=\hat H_{S}+\hat H_{E}+\hat H_{SE}
\label{eq8} 
\ea

and 
\ba
\hat H_{S}|i_{k}\rangle=\epsilon_{i_{k}}|i_{k}\rangle
\notag\\
\hat H_{E}|\gamma\rangle=E_{\gamma}|\gamma\rangle
\label{eq9}
\ea
where $\{|i_{k}\rangle\}$ and $\{|\gamma\rangle\}$ are the eigenvector bases of $\hat H _{S}$ and $\hat H_{E}$. At time $t$ the evolution of $S+E$ is given by

\ba
\hat\rho_{S+E}(t)= \hat U(t)\hat \rho_{S+E}(0) \hat U^{+}(t)
\label{eq10} 
\ea
where $\hat U(t,0)=e^{-i\hat H t}$ is the evolution operator of $S+E$.

%\newpage

\subsection{Explicit expressions of the energy exchange and the speed of the flow}

Using the definitions given in Eqs. (4) and (5) the energy transfer and speed of the energy flow read

\ba
\Delta E(t)=\frac{1}{2}Tr_{S}Tr_{E}\{[\hat H_{E},\hat U(t,0)]\hat\rho_{S+E}(0)\hat U^{+}(t,0)\}+h.c.
\label{eq11} 
\ea
and

\ba
V_{E}(t)=\frac{1}{2}Tr_{S}Tr_{E}\{[\hat H_{E},\frac{d}{dt}\hat U(t,0)]\hat\rho_{S+E}(0)\hat U^{+}(t,0)\}+h.c.
\label{eq12} 
\ea

Developing these expressions in the bases of states given above they read 

\ba
\Delta E(t)=\frac{1}{2}\sum_{j,\gamma}\sum_{i_{1},i_{2},\gamma_{1}}c_{i_{1}} c{*}_{i_{2}}
\langle j \gamma|[\hat H_{E},e^{-i\hat H t}]|i_{1} \gamma_{1}\rangle 
\notag\\
d_{\gamma_{1},\gamma_{1}}\langle i_{2} \gamma_{1}|e^{+i\hat H t}| j \gamma \rangle +h.c.
\label{eq13} 
\ea
and

\ba
V_{E}(t)=\frac{(-i)}{2} \sum_{j,\gamma}  \sum_{\alpha_{1},i_{1},i_{2}}            
c_{i_{1}} c^{*}_{i_{2}} d_{\alpha_{1},\alpha_{1}}
\notag\\
\sum_{j_{1}, \gamma_{1}}\{\langle j \gamma|[\hat H_{E},\hat H_{SE}]|j_{1}, \gamma_{1}\rangle 
\langle j_{1}, \gamma_{1}|e^{-i\hat H t}|i_{1}\alpha_{1}\rangle 
\langle i_{2}\alpha_{1}||e^{+i\hat H t}|j \gamma \rangle 
\notag\\
+ \langle j \gamma|\hat H|j_{1} \gamma_{1}\rangle  
\langle j_{1} \gamma_{1}|[\hat H_{E},e^{-i\hat H t}]|i_{1} \alpha_{1}\rangle 
\langle i_{2} \alpha_{1}|e^{+i\hat H t}|j \gamma \rangle 
\notag\\
- \langle j \gamma|\hat H e^{-i\hat H t}|i_{1} \alpha_{1}\rangle  
\langle i_{2}\alpha_{1}||[\hat H_{E},  e^{+i\hat H t}]|j \gamma \rangle \}
+h.c.
\label{eq14} 
\ea

\section{Properties of the interaction Hamiltonian}

We use now the general expressions of the energy transfer and its exchange rate in order to test the role of the interaction Hamiltonian in this process. Since $S$ and $E$ are distinct different physical systems their Hamiltonians verify the commutation relation $[\hat H_{S},\hat H_{E}]=0$. Former work ~\cite{kr1} has shown that one may consider two cases which are of special interest:\\
 
(a) $[\hat H_{E},\hat H_{SE}]=0$\\ 
 
(b) $[\hat H_{S},\hat H_{SE}]=0$  

\subsection{Case (a)} 
 
It has been shown elsewhere ~\cite{kr1,kr2} that if $H_{E}$ and $H_{SE}$ commute the evolution of the system $S$ is characterized by the divisibility property which is a specific property of Markovian systems. 
Since the Hamiltonian of the system $S$  commutes in practice with $\hat H_{E}$ it follows that $\hat H_{E}$ commutes with the whole Hamiltonian $\hat H$, hence also $\hat U(t,0)$ and its derivatives. Going back to the expressions of $\Delta E(t)$ and $V_{E}(t)$ it comes out that $\Delta E(t)=V_{E}(t)=0$.

There is no energy exchange between $S$ and $E$ in this case. The physical explanation is the following: the divisibility property imposes that the environment stays in a fixed state at a fixed energy which blocks any possible transfer of energy between the two parts of the total system $S+E$. This is due to the fact that $\hat H_{SE}$ is diagonal in $E$ space, hence the considered state $|\gamma \rangle$ stays the same over any time interval.
 
\subsection{Evolution of the energy: an impurity immersed in a bosonic condensate}
 
We introduce the Hamiltonian of a fermionic impurity interacting with a Bose-Einstein condensate ~\cite{chr}. It is given by $\hat H= \hat H_{S}+\hat H_{E}+\hat H_{SE}$ where

\begin{center}
\ba
\hat H_{S}=\sum_{\vec k}\epsilon_{\vec k}c^{+}_{\vec k}c_{\vec k}
\label{eq15} 
\ea
\end{center}
\begin{center} 
\ba
\hat H_{E}=\sum_{\vec k}e_{\vec k}a^{+}_{\vec k}a_{\vec k}+\frac{1}{2V}\sum_{\vec k_{1}\vec k_{2}
\vec q}V_{B}(\vec q)(\vec q)a^{+}_{\vec k_{1}}a^{+}_{\vec k_{1}+\vec q}a^{+}_{\vec k_{2}-\vec q}
a_{\vec k_{2}}a_{\vec k_{1}}
\label{eq16} 
\ea
\end{center}
\begin{center}
\ba
\hat H_{SE}=\frac{1}{V}\sum_{\vec k_{3}\vec k_{4}\vec q}c^{+}_{\vec k_{3}+\vec q}c_{\vec k_{4}}
a^{+}_{\vec k_{4}-\vec q}a_{\vec k_{3}}
\label{eq17} 
\ea
\end{center}
where $[c,c^{+}]$ and $[a,a^{+}]$ are fermion and boson annihilation and creation operators. 

We consider the case where the momentum transfer $\vec q=0$. Then one expects that there is no energy exchange between $S$ and $E$. This is indeed so since a simple calculation shows that 
$[\hat H_{E},\hat H_{SE}]=0$ in this case.  
 
\subsection{Case (b)}
 
Start from the general expression of $\Delta E(t)$ given by Eq.(13). 

Since $[\hat H_{S},\hat H_{SE}]=0$ all the matrix elements are diagonal in $S$ space and the expression takes the form

\ba
\Delta E(t)=\sum_{j}|c_{j}|^{2}\sum_{\gamma,\gamma_{1}}(E_{\gamma}-E_{\gamma_{1}})
\langle j \gamma|e^{-it \hat H}|j \gamma_{1}\rangle d_{\gamma_{1}\gamma_{1}}
\langle j \gamma_{1}|e^{+it \hat H}|j \gamma \rangle
\label{eq18}  
\ea

In order to follow the evolution of $\Delta E(t)$ in time we determine the expression of $V_{E}(t)$.
A somewhat lengthy but straightforward calculation leads to the following expression:

\ba
V_{E}(t)=V^{(1)}_{E}(t)+ c.c. +V^{(2)}_{E}(t)+ c.c.
\label{eq19}
\ea
where 
\ba
V^{(1)}_{E}(t)=(-i)\sum_{j}|c_{j}|^{2}\sum_{\gamma,\gamma_{1}}(E_{\gamma}-E_{\gamma_{1}})
\langle j \gamma|\hat H_{SE}|j \gamma_{1}\rangle
\notag\\
\sum_{\gamma_{2}}\langle j \gamma_{1}|e^{-it \hat H}|j \gamma_{2}\rangle 
d_{\gamma_{2},\gamma_{2}} \langle j \gamma_{2}|e^{+it \hat H}|j \gamma \rangle  
\label{eq20}  
\ea
and $V^{(1)*}_{E}(t)$ its complex conjugate. It comes out that the c.c. $V^{(1)*}_{E}(t)=V^{(1)}_{E}(t)$ which means that $V^{(1)}_{E}(t)$ is real. 
The second term reads

\ba 
V_{E}^{(2)}(t)= (-i)\sum_{j}|c_{j}|^{2}\sum_{\gamma \gamma_{1}}(E_{\gamma}+\epsilon_{j})
\sum_{\gamma_{2}}(E_{\gamma}-E_{\gamma_{2}})   
\notag\\
\langle j \gamma_{1}|e^{-it \hat H}|j \gamma_{2}\rangle 
d_{\gamma_{2},\gamma_{2}}\langle j \gamma_{2}|e^{+it \hat H}|j \gamma \rangle 
\label{eq21}  
\ea 

It is easy to see that $V_{E}^{(2)}(t) +c.c.=0$, hence $V_{E}(t)=2V_{E}^{(1)}(t)$.

For $t=0$ 

\ba
\Delta E(0)= \sum_{j}|c_{j}|^{2}\sum_{\gamma,\gamma_{1}}(E_{\gamma}-E_{\gamma_{1}})
\langle j \gamma|j \gamma_{1}\rangle d_{\gamma_{1}\gamma_{1}}
\langle j \gamma_{1}|j \gamma \rangle=
\notag\\
\sum_{j}|c_{j}|^{2}\sum_{\gamma,\gamma_{1}}(E_{\gamma}-E_{\gamma_{1}})\delta_{\gamma \gamma_{1}}
\label{eq22}  
\ea
Hence $\Delta E(0)=0$ which could have been anticipated from the symmetry property of the expression of $\Delta E(t)$.\\

But contrary to case $(a)$ the energy tranfer is now different from zero. It varies with time hence $\Delta E(t)$ increases or de decreases as a function of the sign of $V_{E}(t)$. This is due to the fact that now many channels can open in $E$ space because $\hat H_{SE}$is no longer diagonal in this space.

\subsection{Evolution of the energy: two examples}

In order to illustrate this case we develop two models on which we exemplify the time dependence of the energy transfer between a system and its environment when $[\hat H_{S},\hat H_{SE}]=0$.

\begin{itemize}

\item {\bf First example}: we consider a model consisting of a 2-level state system $E$, 
$[|\gamma\rangle=|1\rangle,|2\rangle]$. The Hamiltonian $\hat H$ decomposes into two parts 
$\hat H_{0}=\hat H_{S}+\hat H_{SE}$ and $\hat H_{E}$. We consider the case where
\ba 
[\hat H_{0},[\hat H_{0},\hat H_{E}]]=[\hat H_{E},[\hat H_{0}, \hat H_{E}]]=0
\label{eq23}  
\ea 
and $[\hat H_{0},\hat H_{E}]=c1$ where $c$ is a number. Then             

\ba 
e^{\hat H_{0}+\hat H_{E}}=e^{\hat H_{0}}e^{\hat H_{E}}e^{c/2}
\label{eq24}  
\ea   
and
   
\ba 
e^{i(\hat H_{0}+\hat H_{E)}t}=e^{i\hat H_{0}t}e^{i\hat H_{E}t}e^{ct^{2}/2}
\label{eq25}  
\ea 

The quantities which enter the expressions which follow are defined in Appendix A.

The expressions of $\Delta E(t)$ and $V_{E}(t)$ read

\ba
\Delta E(t)=e^{ct^{2}}\Delta_{12}\sum_{j}|c_{j}|^{2}(d_{22}-d_{11})[a_{j}^{(12)2}(t)+b_{j}^{(12)2}(t)]
\label{eq26}  
\ea 
where $\Delta_{12}=E_{1}-E_{2}$, $d_{11},d_{22}$ the weights of the states in $E$ space and

\ba
V_{E}(t)=2e^{ct^{2}} \Delta_{12}\sum_{j} |c_{j}|^{2}[I^{(j)}_{12}Re(\langle 2|\hat \Omega_{j}t)|1\rangle + R^{(j)}_{12}
Im (\langle 2|\hat \Omega_{j}(t)|1\rangle]                                
\label{eq27}  
\ea 
with 

\ba
Re\langle 2|\hat \Omega_{j}(t)|1\rangle=a_{j}^{11}(t)[a_{j}^{21}(t)\cos(\Delta_{12}t)+b_{j}^{21}(t)
\sin(\Delta_{12}t)]d_{11}
\notag\\
+a_{j}^{22}(t)a^{21}(t)d_{22}
\notag\\
Im\langle 2|\hat \Omega_{j}(t)|1\rangle=a_{j}{11}(t)
[-b_{j}^{21}(t)\cos(\Delta_{12}t)+a_{j}^{21}(t)\sin(\Delta_{12}t)]d_{11}
\notag\\
+b_{j}^{21}(t)a_{j}^{22}(t)d_{22}
\label{eq28} 
\ea 
 
Both $\Delta E(t)$ and $V_{E}(t$ are oscillating functions of time. $\Delta E(t)$ keeps a fixed sign depending on the sign of $\Delta_{12}$, $V_{E}(t)$ may change sign with time. The energy and the speed of the energy transfer decays to zero for real $c$ real and negative.\\
 
\item 
{\bf Second example}: we consider the Hamiltonian $\hat H=\hat H_{S}+\hat H_{E}+\hat H_{SE}$ which governs the coupling of a phonon field to the electron in the BCS theory of superconductivity.

The total Hamiltonian of the electron-phonon system reads 

\begin{center}
\ba
\hat H_{S}=\sum_{\vec k_{1}}\epsilon_{\vec k_{1}}c^{+}_{\vec k_{1}}c_{\vec k_{1}}
\label{eq29} 
\ea
\end{center}
\begin{center} 
\ba
\hat H_{E}=\sum_{\vec q}\hbar \omega_{\vec q}a^{+}_{\vec q}a_{\vec q}
\label{eq30} 
\ea
\end{center}
\begin{center}
\ba
\hat H_{SE}=V_{ph-e}(\vec q)\sum_{\vec k_{2}\vec q}(a^{+}_{-\vec q}+a_{\vec q})
c^{+}_{\vec k_{2}+\vec q}c_{\vec k_{2}}
\label{eq31} 
\ea
\end{center}
where $V_{ph-e}(\vec q)$ is the phonon-electron interaction, $[a,a^{+}]$ and $[c,c^{+}]$ are phonon and electron annihilation and creation operators. We consider the case where the phonons evolve in the zero mode $\vec q=0$.

Then $\hat H_{E}=\hbar \omega_{0}a^{+}_{0}a_{0}$ and 
$\hat H_{SE}=V_{ph-e}(0)\sum_{\vec k_{2}}(a^{+}_{0}+a_{0})c^{+}_{\vec k_{2}}c_{\vec k_{2}}$.

The density operator of the total system $S+E$ at $t=0$ is given by the expression 
\ba
\hat \rho^{(n_{0},n'_{0})}_{(\vec k,\vec k')}=\frac{1}{2\pi(n_{0}!n'_{0}!)^{1/2}}|\vec k n_{0}\rangle    \langle \vec k' n'_{0}|
\label{eq32}  
\ea 
Working out the commutation relation between $\hat H_{S}$ and $\hat H_{SE}$ leads to 
$[\hat H_{S},\hat H_{SE}]=0$, hence the evolution operator can be written as 
\ba
e^{-i\hat Ht}=e^{-i\hat H_{S}t}e^{-i(\hat H_{E}+\hat H_{SE})t}
\label{eq33}  
\ea 
and in explicit form

\ba
e^{-i\hat Ht}=e^{-it\sum_{k=1}^{\nu}\epsilon_{\vec k}c^{+}_{\vec k}c_{\vec k}}
e^{-i[\omega_{0} a^{+}_{0}a_{0}+\nu V_{ph-e}(0)(a^{+}_{0}+a_{0})]t}                                         
\label{eq34}  
\ea
where $\nu$ is the number of electron states. The quantity of interest concerns the time dependence of $\Delta E(t)$ and $V_{E}(t)$ given by Eqs.(18) and (19), hence the time evolution of the matrix elements of $e^{-i\hat Ht}$ 

\ba
E^{(n_{0},n'_{0})}_{\vec k \vec k}(t)=\langle \vec k n_{0}|e^{-it\sum_{k=1}^{\nu}\epsilon_{\vec k}c^{+}_{\vec k}c_{\vec k}}e^{-i[\omega_{0} a^{+}_{0}a_{0}+\nu V_{ph-e}(0)(a^{+}_{0}+a_{0})]t}|\vec k n'_{0}\rangle    
\label{eq35}  
\ea
These matrix elements and their hermitic conjugates which enter the expressions of $\Delta E(t)$ and $V_{E}(t)$ can be worked out explicitly using the Zassenhaus development~\cite{za}, see Appendix B and~\cite{ca}.

They read
\ba
E^{(n_{0},n'_{0})}_{\vec k \vec k}(t)=N^{-1}e^{-i\epsilon_{k}t}e^{-i\omega_{0}n_{0}t}  F^{(n_{0},n'_{0})}(t)  
\label{36}
\ea
with
\ba
F^{(n_{0},n'_{0})}(t)= \sum_{n_{2}\leq n_{0},n_{2}\leq n_{3}}
\sum_{n_{4}\leq n_{3},n_{4}\leq n'_{0}}(-i)^{n_{0}+n_{3}}(-1)^{n'_{0}+n_{2}-n_{4}}
\notag\\
\frac{n_{0}!n'_{0}!(n_{2}!)^{2}(n_{3}!)^{2}[\alpha(t)^{n_{0}+n_{3}-2n_{2}}]
[\zeta(t)^{n'_{0}+n_{3}-2n_{4}}]}{(n_{2})^{2} (n_{4})^{2}(n_{0}-n_{2})!(n_{3}-n_{4})!
(n_{3}-n_{2})!(n'_{0}-n_{4})!}e^{\Psi(t)}
\label{37} 
\ea
and the normalization factor $N= 2\pi(n_{0}!n'_{0}!)^{1/2}$. The functions $\alpha(t)$, $\zeta(t)$
and $\Psi(t)$ read

\ba
\alpha(t)=\frac{\nu V_{ph-e}(0)\sin\omega_{0}t}{\omega_{0}}
\label{eq38}      
\ea

\ba
\zeta(t)=\frac{\omega_{0}[1-\cos\nu V_{ph-e}(0)t]}{\nu V_{ph-e}(0)}
\label{eq39}      
\ea

\ba
\Psi(t)=-\frac{1}{2}[\frac{\nu^{2}V^{2}_{ph-e}(0)\sin^{2}(\omega_{0} t)}{\omega_{0}^{2}}+\frac{\omega_{0}^{2}(1-\cos\nu V_{ph-e}(0)t)^{2}}{\nu^{2}V_{ph-e}^{2}}]                           
\label{eq40}      
\ea  

As one can see from these expressions $E^{(n_{0},n'_{0})}_{\vec k \vec k}(t)$ are oscillating functions of time which leads to the conclusion that the energy transfer and the transfer velocity oscillate continuously and stay finite over any interval of time. 

\end{itemize} 
 
\section{Conclusions, remarks}

In the present work we used a cumulant approach~\cite{gua} in order to study the energy transfer between a system and its environment and the rate at which this transfer evolves in time. 

If the Hamiltonian of the environment commutes with the interaction between the system and the environment 
there results an absence of energy transfer between the two subsystems. This can be explained in the following way. As already seen in former work the commutation property is a sufficient condition for divisibility in the time behaviour of an open system, one of the properties which characterize Markov processes ~\cite{sti,riv}. 
In this case it has been shown~\cite{kr1,kr2,kr3} that the environment keeps in the energy state in which it was at the origin of time and stays there over any interval of time. Hence the energy of the environment is blocked in a fixed state so that it cannot feed the system and does not receive energy from it. Time delays are correlated with the possibility of the environment to jump between different states in closed systems as well as in open ones (see f.i.~\cite{dod,def1,def2} and references quoted in there). This is the case  in non-Markovian systems ~\cite{zhe,hai,ren,gua1}. 

The experimental realization of the absence of energy transfer may be obtained under different conditions:

\begin{itemize}

\item the strength of the interaction can be chosen such that it keeps very weak and hence does not allow any possible jump to another level in the case of a discrete environment spectrum.

\item the temperature of the environment is kept close to zero so that the ground state is the only accessible state.

\item the commutation relation between the environment and the interaction is rigorously verified which is the case discussed in the present work.

\end{itemize}

We considered also the case where the system and the interaction Hamiltonians commute with each other.
In this case energy can flow from the environment to the system and back, there is no blocking effect
coming from the environment since several states are at hand and the energy exchange can take place.
In order to illustrate this situation we worked out two model systems corresponding to different physical situations. In the first case the energy exchange oscillates but stops exponentially, in the second case which models the electron-phonon system the energy exchange and the exchange speed oscillate in time and never go to zero. This is so because the system behaves coherently as it has been shown in ref.~\cite{kr3,lid}.

\section{Appendix A}    

We define the following matrix elements which enter the expressions of $\Delta E(t)$ and $V_{E}(t)$ given by Eqs.(16-18)
   
\ba 
\langle j 1|e^{-i\hat Ht}|j 1\rangle=e^{ct^{2}/2}e^{-i\epsilon_{j}t}e^{-iE_{1}t}a_{j}^{11}(t)
\notag\\
\langle j 2|e^{+i\hat Ht}|j 2\rangle=e^{ct^{2}/2}e^{+i\epsilon_{j}t}e^{+iE_{2}t}a_{j}^{22}(t)
\notag\\
\langle j 1|e^{+i\hat Ht}|j 2\rangle=e^{ct^{2}/2}e^{+i\epsilon_{j}t}e^{+iE_{1}t}
(a_{j}^{12}(t)+ib_{j}^{12}(t))
\notag\\
\langle j 2|e^{-i\hat Ht}|j 1\rangle=e^{ct^{2}/2}e^{-i\epsilon_{j}t}e^{-iE_{2}t}
(a_{j}^{21}(t)-ib_{j}^{21}(t))
\label{eq41}  
\ea    
where $\epsilon_{j}$ is the eigenvalue of state $|j\rangle$ and $|E_{k}\rangle,(k=1,2)$ the eigenvalues of states the states $|\gamma\rangle$
\ba 
a_{j}^{11}(t)=Re(\langle j 1|e^{-i \hat H_{SE}t}|j 1 \rangle)
\notag\\
a_{j}^{22}(t)=Re(\langle j 2|e^{+i \hat H_{SE}t}|j 2 \rangle)
\notag\\
a_{j}^{12}(t)=Re(\langle j 1|e^{+i \hat H_{SE}t}|j 2 \rangle)
\notag\\
a_{j}^{21}(t)=Re(\langle j 2|e^{-i \hat H_{SE}t}|j 1 \rangle)
\notag\\
b_{j}^{12}(t)=Im(\langle j 1|e^{+i \hat H_{SE}t}|j 2 \rangle)
\notag\\
b_{j}^{21}(t)=Im(\langle j 2|e^{-i \hat H_{SE}t}|j 1 \rangle)
\label{eq42}  
\ea
Introduce also

\ba
R^{(j)}_{12}=Re(\langle j 1|\hat H_{SE}|j 2 \rangle)
\notag\\
I^{(j)}_{12}=Im(\langle j 1|\hat H_{SE}|j 2 \rangle)  
\label{eq43}
\ea  
and

\ba 
\hat \Omega_{j}(t)=\sum_{\gamma}e^{-i\hat H t}|j \gamma\rangle d_{\gamma \gamma}\langle j \gamma|
e^{+i\hat H t}
\label{eq44}  
\ea 

Using these defintions the calculation leads to expressions (26) and (27) of $\Delta E(t)$ and $V_{E}(t)$ in the text.

\section{Appendix B: the Zassenhaus development} 
   
If 
$X=-i(t-t_{0})(\hat H_{S}+\hat H_{E})$ and $Y=-i(t-t_{0})\hat H_{SE}$

\ba
e^{X+Y}=e^{X}\otimes e^{Y}\otimes e^{-c_{2}(X,Y)/2!}\otimes e^{-c_{3}(X,Y)/3!}\otimes e^{-c_{4}(X,Y)/4!}...
\label{eq45}
\ea
where

\begin{center}
$c_{2}(X,Y)=[X,Y]$\\ 
$c_{3}(X,Y)=2[[X,Y],Y]+[[X,Y],X]$\\ 
$c_{4}(X,Y)=c_{3}(X,Y)+3[[[X,Y],Y],Y]+[[[X,Y],X],Y]+[[X,Y],[X,Y]$\\
\end{center} 
 
The series has an infinite number of term which can be generated iteratively in a straightforward way ~\cite{ca}. If $[X,Y]=0$ the truncation at the  third term  leads to the factorisation of the $X$ and the $Y$ contribution. If $[X,Y]=c$ where $c$ is a c-number the expression corresponds to the well-known Baker-Campbell-Hausdorff formula.

\end{document}